\def\rfr#1{eq. (\ref{#1})}

\def\Rfr#1{Eq. (\ref{#1})}

\def\bm#1{{\mbox{\boldmath$#1$\unboldmath}}}
\def\asec{$''$ cy$^{-1}$}

\def\asec{$''$ cy$^{-1}$}

\def\bar{\begin{eqnarray}}
\def\ear{\end{eqnarray}}

\def\eqi{\begin{linenomath*}\begin{equation}}
\def\eqf{\end{equation}\end{linenomath*}}
\def\eqia{\begin{eqnarray}}
\def\eqfa{\end{eqnarray}}
\def\rp#1#2{{#1\over#2}}

\def\lb#1{\label{#1}}

\def\oc2{$\mathcal{O}(c^{-2})$}

\documentclass[11pt]{article}
\usepackage{amsmath,amsthm,amscd,amssymb}
\usepackage{latexsym}
\usepackage{graphicx,epsfig}
\usepackage[edtable]{lineno}

\begin{document}

\noindent{\bf \LARGE{First preliminary tests of the general
relativistic gravitomagnetic field of the Sun and new constraints
on a Yukawa-like fifth force from planetary data}}
\\
\\
\\
{L. Iorio, }\\
{\it Viale Unit$\grave{a}$ di Italia 68, 70125\\Bari, Italy
\\tel./fax 0039 080 5443144
\\e-mail: lorenzo.iorio@libero.it}

\begin{abstract}
The general relativistic  Lense-Thirring precessions of the
perihelia of the inner planets of the Solar System are $\lesssim
10^{-3}$ arcseconds per century.  Recent improvements in
planetary orbit determination may yield the first
observational evidence of such a tiny effect. Indeed,
corrections to the known perihelion rates of $-0.0036\pm
0.0050$, $-0.0002\pm 0.0004$ and $0.0001\pm 0.0005$ arcseconds per
century were recently estimated by E.V. Pitjeva for Mercury, the
Earth and Mars, respectively, on the  basis of the EPM2004
ephemerides and a set of more than 317,000 observations of various
kinds. The predicted relativistic Lense-Thirring precessions for
these planets are $-0.0020$, $-0.0001$ and $-3\times 10^{-5} $
arcseconds per century, respectively and are compatible with the
determined perihelia corrections.  The relativistic predictions
fit better than the zero-effect hypothesis, especially if a
suitable linear combination of the perihelia of Mercury and the
Earth, which a priori cancels out any possible bias due to the
solar quadrupole mass moment, is considered. However, the
experimental errors are still  large. Also the latest data for
Mercury  processed independently by Fienga et al. with the INPOP
ephemerides yield preliminary insights about the existence of the
solar Lense-Thirring effect. The data from the forthcoming
planetary mission BepiColombo will
improve our knowledge of the orbital motion of this planet and,
consequently, the precision of the measurement of the
Lense-Thirring effect. As a by-product of the present analysis, it
is also possible to constrain the strength of a Yukawa-like fifth
force to a $10^{-12}-10^{-13}$ level at scales of about one
Astronomical Unit ($10^{11}$ m).
\end{abstract}



\section{Introduction}
A satisfactorily empirical corroboration of a fundamental theory
requires that as many independent experiments as possible are
conducted by different scientists in different laboratories. Now,
the general relativistic gravitomagnetic Lense-Thirring (LT)
effect is difficult to test, also in the weak-field and
slow-motion approximation, valid in our Solar System, both because
such a relativistic effect is very small and the competing
classical signals are often quite larger. Until now, a $6\%$ LT
test has recently been conducted in the gravitational field of
Mars by using the data of the Mars Global Surveyor (MGS)
spacecraft; other tests accurate to about\footnote{Other estimates
point towards a $15-20\%$ error.} $10\%$ have been performed by a
different team in the gravitational field of the Earth by
analyzing the data of the LAGEOS and LAGEOS II artificial
satellites (see Section \ref{pppo}).  We think it is worthwhile to
further extend these efforts trying  to use different
laboratories, i.e. other gravitational fields, even if the
outcomes of such tests should be less accurate than those
conducted so far. To this aim, in this paper we investigate the LT
effect induced by the Sun on the orbital motion of the inner
planets of the Solar System in the context of the latest results
in the planetary ephemerides field.

\subsection{The Lense-Thirring effect}
According to Einstein, the action of the gravitational potential
$U$ of a given distribution of mass-energy is described by the
coefficients $g_{\mu\nu},\ \mu,\nu=0,1,2,3$, of the space-time
metric tensor. They are determined, in principle, by solving the
fully non-linear field equations of the Einsteinian General Theory
of Relativity (GTR) for the considered mass-energy content. These
equations can be linearized in the weak-field ($U/c^2 <<1$, where
$c$ is the speed of light in vacuum) and slow-motion ($v/c<<1$)
approximation (Mashhoon 2001; Ruggiero and Tartaglia 2002), valid
throughout the Solar System, and look like the equations of the
linear Maxwellian electromagnetism. Among other things, a
noncentral, Lorentz-like force
\eqi\bm F_{\rm LT}=-2m\left(\rp{\bm v}{c}\right)\times \bm
B_g\lb{fgm}\eqf
%
acts on a moving test particle of mass $m$. It  is
induced by the post-Newtonian component $\bm B_g$ of the
gravitational field in which the particle moves with velocity \bm
v. $\bm B_g$ is related to the mass currents of the mass-energy
distribution of the source and comes from the off-diagonal
components $g_{0i}, i=1,2,3$ of the metric tensor. Thanks to such
an analogy, the ensemble of the gravitational effects induced by
mass displacements is also named gravitomagnetism. For a central
rotating body of mass $M$ and proper angular momentum \bm L the
gravitomagnetic field is\eqi\bm B_g=\rp{G[3\bm r(\bm r\cdot \bm L
)-r^2 \bm L]}{cr^5}.\lb{gmfield}\eqf

One of the consequences of \rfr{fgm} and \rfr{gmfield} is a
gravitational spin--orbit coupling. Indeed, if we consider the
orbital motion of a particle in the gravitational field of a
central spinning mass,  it turns out that
%
the longitude of the ascending node $\Omega$ and the argument of
pericentre $\omega$ of the orbit of the test particle are affected
by tiny secular advances $\dot\Omega_{\rm LT}$, $\dot\omega_{\rm
LT}$ (Lense and Thirring 1918, Barker and O'Connell 1974, Cugusi
and Proverbio 1978, Soffel 1989, Ashby and Allison 1993, Iorio
2001) \eqi \dot\Omega_{\rm LT} =\frac{2GL}{c^2
a^3(1-e^2)^{\frac{3}{2}}},\ \dot\omega_{\rm LT} =-\frac{6GL\cos
i}{c^2 a^3(1-e^2)^{\frac{3}{2}}},\lb{LET} \eqf where $a,\ e$ and
$i$ are the semimajor axis, the eccentricity and the inclination,
respectively, of the orbit and $G$ is the Newtonian gravitational
constant. Note that in their original paper Lense and Thirring
(1918) used the longitude of pericentre
$\varpi\equiv\Omega+\omega$.

The gravitomagnetic force may have strong consequences in many
astrophysical and astronomical scenarios involving, e.g.,
accreting disks around black holes (Thorne et al. 1986; Stella et al. 2003),
gravitational lensing and time delay (Sereno 2003; 2005a; 2005b).
Unfortunately, in these contexts the knowledge of the various
competing effects is rather poor and makes very difficult to
reliably extract the genuine gravitomagnetic signal from the noisy
background. E.g., attempts to measure the LT effect around black
holes are often confounded by the complexities of the dynamics of
the hot gas in their accretion disks. On the contrary, in the
solar and terrestrial space environments the LT effect is weaker
but the various sources of systematic errors are relatively well
known and we have the possibility of using various artificial and
natural orbiters both to improve our knowledge of such biases and
to design suitable observables circumventing these problems, at
least to a certain extent.

\subsection{The performed and ongoing tests}\lb{pppo}
Up to now, all the performed and ongoing tests of gravitomagnetism
were implemented in the weak-field and slow-motion scenarios of
the Earth and Mars gravitational fields.

As far as the Earth is concerned, in April 2004 the GP-B
spacecraft (Everitt 1974; Fitch et al. 1995; Everitt et al. 2001) was launched. Its
aim is the measurement of another gravitomagnetic effect, i.e. the
precession of the spins (Pugh 1959; Schiff 1960) of four
superconducting gyroscopes carried onboard. The level of accuracy obtained so far is about
$256-128\%$ (Muhlfelder et al. 2007), with the hope of reaching\footnote{See on the WEB StanfordNews 4/14/07 downloadable at http://einstein.stanford.edu/.}  $\approx 13\%$ in December 2007.

In regard to the LT effect on the orbit
of a test particle, the idea of using
the LAGEOS satellite and, more generally, the Satellite Laser
Ranging (SLR) technique to measure it in
the terrestrial gravitational field with the existing artificial
satellites was put forth for the first time by Cugusi and
Proverbio (1978). Attempts to practically implement such a strategy
began in 1996 with the LAGEOS and LAGEOS II satellites (Ciufolini et al. 1996).
The latest test was performed by Ciufolini and Pavlis (2004).
They analyzed the data of  LAGEOS and LAGEOS II by using an
observable independently proposed by Pavlis (2002), Ries et al. (2003a, 2003b)
and Iorio and Morea (2004). The error claimed by Ciufolini and
Pavlis (2004) is 5-10$\%$ at 1-3 sigma, respectively. The
assessment of the total accuracy of such a test raised a
debate (Iorio 2005a; 2005b; 2006a; 2007; Ciufolini and Pavlis
2005; Lucchesi 2005).

Recently, a $6\%$ LT test on the orbit of the Mars Global Surveyor
(MGS) spacecraft in the gravitational field of Mars has been
reported (Iorio 2006b); indeed, the predictions of general
relativity are able to accommodate, on average, about $94\%$ of
the measured residuals in the out-of-plane part of the MGS orbit
over 5 years.

Finally, it must be noted that, according to Nordtvedt (2003), the
multi-decade analysis of the Moon'orbit by means of the Lunar
Laser Ranging (LLR) technique yields a comprehensive test of the
various parts of order $\mathcal{O}(c^{-2})$ of the post-Newtonian
equation of motion. The existence of the gravitomagnetic interaction as predicted
by GTR would, then, be inferred from the high accuracy
of the lunar orbital reconstruction.  A 0.1$\%$ test was recently reported (Murphy et al. 2007): a critical discussion of the real sensitivity of LLR to gravitomagnetism can be found in (Kopeikin 2007).
Also the radial motion of the
LAGEOS satellite would yield another indirect confirmation of the
existence of the gravitomagnetic interaction (Nordtvedt 1988).

\section{The solar gravitomagnetic field}
The action of the solar gravitomagnetic field on the Mercury's
longitude of perihelion was calculated for the first time by de
Sitter (1916) who, by assuming an homogenous and uniformly
rotating Sun, found a secular advance of $-0.01$ arcseconds per
century ( \asec\ in the following). This value is also quoted by
Soffel (1989). Cugusi and Proverbio (1978) yield $-0.02$ \asec\
for the argument of perihelion of Mercury. Instead, recent
determinations of the Sun's proper angular momentum
$L_{\odot}=(190.0\pm 1.5)\times 10^{39}$ kg m$^2$ s$^{-1}$ from
helioseismology (Pijpers 1998; 2003), accurate to $0.8\%$, yield a
precessional effect one order of magnitude smaller for Mercury
(Ciufolini and Wheeler 1995; Iorio 2005c). See Table \ref{solgm}
for the gravitomagnetic precessions of the four inner planets.
{\small\begin{table}\caption{Gravitomagnetic secular precessions
of the longitudes of  perihelion $\varpi$ of Mercury, Venus, Earth
and Mars in \asec. The value $(190.0\pm 1.5)\times 10^{39}$ kg
m$^2$ s$^{-1}$ (Pijpers 1998; 2003) has been adopted for the solar
proper angular momentum $L_{\odot}$. }\label{solgm}

\begin{tabular}{llll}
\noalign{\hrule height 1.5pt}

 Mercury & Venus  & Earth & Mars\\
-0.0020 & -0.0003 & -0.0001 & $-3\times 10^{-5}$\\ \hline

\noalign{\hrule height 1.5pt}
\end{tabular}

\end{table}}
 As can be seen, they are of the order of $10^{-3}-10^{-5}$
\asec.

So far, the LT effect on the orbits of the Sun's planets was
believed to be too small to be detected (Soffel 1989). Iorio
(2005c) preliminarily investigated the possibility of measuring
such tiny effects in view of recent important developments in the
planetary ephemerides generation. It is remarkable to note that
the currently available estimate of $L_{\odot}$ is accurate enough
to allow, in principle, a genuine test of GTR. Moreover, it was
determined in a relativity-free fashion from astrophysical
techniques which do not rely on the dynamics of planets in the
gravitational field of the Sun. Thus, there is no any a priori
`memory' effect of GTR itself in the adopted value of $L_{\odot}$.
\section{Compatibility of the determined extra-precessions of planetary perihelia with
the LT effect}
\subsection{The Keplerian orbital elements}
The Keplerian orbital elements like $\varpi$ are not directly
observable quantities like right ascensions, declinations, ranges
and range-rates which can be measured from optical observations,
radiometric measurements, meridian transits, etc. They can only be
computed from a state vector in rectangular Cartesian coordinates
which also allows to compute predicted values of the observations.
In this sense, speaking of an ``observed'' time series of a
certain Keplerian element would mean that it has been computed
from the machinery of the data reduction of the real
observations\footnote{In practice, this procedure cannot be always
performed at all times because the observations collected at a
given time not always allow for the computation of the entire
state vector in rectangular Cartesian coordinates.
}. Keeping this in mind, it would be possible, in principle, to
extract the LT signal from the planetary motions by taking the
difference between two suitably computed time-series of the
Keplerian elements in such a way that it fully accounts for the
gravitomagnetic signature. Such ephemerides, which should share
the same initial conditions, would differ in the fact that one
would be based on the processing of the real data, which are
presumed to fully contain also the LT signal, and the other one
would, instead, be the result of a purely numerical propagation.
The dynamical force models with which the data are to be processed
and the numerical ephemeris  propagated do not contain the
gravitomagnetic force itself: only the general relativistic
gravitoelectric terms must be present. Moreover, the astronomical
parameters entering the perturbations which can mimic the LT
signature should not be fitted in the data reduction process: they
should be kept fixed to some reference values, preferably obtained
in a relativity-independent way so to avoid `imprinting' effects.
Thus, in the resulting ``residual'' time series $\Delta\varpi_{\rm
obs}(t)$, the LT signature should be entirely present.
\subsection{The EPM2004 ephemerides}
A somewhat analogous procedure was recently  implemented with the
Ephemerides of Planets and the Moon EPM2004 (Pitjeva 2005a;
2005b) produced by the Institute of Applied Astronomy (IAA) of the Russian Academy of Sciences (RAS).
They are based on a data set of more than 317,000
observations (1913-2003) including radiometric measurements of
planets and spacecraft, astrometric CCD observations of the outer
planets and their satellites, and meridian and photographic
observations. Such ephemerides were constructed by the
simultaneous numerical integration of the equations of motion for
all planets, the Sun, the Moon, 301 largest asteroids, rotations
of the Earth and the Moon, including the perturbations from the
solar quadrupolar mass moment $J^{\odot}_2$ and asteroid ring that
lies in the ecliptic plane and consists of the remaining smaller
asteroids. In regard to the post-Newtonian dynamics, only the
gravitoelectric terms, in the harmonic gauge, were included
(Newhall et al. 1983).
\subsection{The measured extra-precessions of the planetary perihelia and the Lense-Thirring effect}
As a preliminary outlook on the measurability  of the
Lense-Thirring perihelion precessions, let us make the following
considerations. The magnitude of the gravitomagnetic shift of the
Mercury's perihelion over a 90-years time span like that covered
by the EPM2004 data amounts to 0.0018 $''$. The accuracy in
determining the secular motion of Mercury's perihelion can be
inferred from the results for the components of the eccentricity
vector $k=e\cos\varpi$ and $h=e\sin\varpi$ reported in Table 4 by
Pitjeva (2005b). Indeed, the formal standard deviations of $k$ and
$h$ are 0.123 and 0.099 milliarcseconds, respectively. Thus, the
formal error in measuring $\varpi$ is about 0.0007 $''$. An
analogous calculation for the Earth yields an error in $\varpi$ of
$8\times 10^{-5}$ $''$.

The EPM2004 ephemerides were used to determine corrections
$\Delta\dot\varpi_{\rm obs}$ to the secular precessions of the
longitudes of perihelia of the inner planets as fitted parameters
of a particular solution. In Table 3 by Pitjeva (2005a), part of
which is reproduced in Table \ref{pitabl}, it is possible to find
their values
 obtained by comparing the model
observations computed using the constructed ephemerides with
actual observations. Note that in determining such
extra-precessions the PPN parameters (Will 1993) $\gamma$ and
$\beta$ and the solar even zonal harmonic coefficient
$J_2^{\odot}$ were not fitted; they were held fixed to their GTR
values, i.e. $\gamma=\beta=1$, and to
 $J_2^{\odot}=2\times 10^{-7}$.
Note also that the unit values of $\beta$ and $\gamma$ were
measured in a variety of approaches which are independent of the
gravitomagentic force itself.
%
{\small\begin{table}\caption{ Determined extra-precessions
$\Delta\dot\varpi_{\rm obs}$ of the longitudes of  perihelia of
the inner planets, in \asec, by using EPM2004  with
$\beta=\gamma=1$, $J_2^{\odot}=2\times 10^{-7}$. The
gravitomagnetic force was not included in the adopted dynamical
force models. Data taken from Table 3 of (Pitjeva 2005a). It is important to note that the quoted
uncertainties are not the mere formal, statistical errors but are realistic in the sense that they
were obtained from comparison of many different
solutions with different sets of parameters and observations (Pitjeva, private communication 2005a). The correlations among such determined planetary perihelia rates are very low with a maximum of about $20\%$ between Mercury and the Earth (Pitjeva, private communication 2005b).
}\label{pitabl}

\begin{tabular}{llll} \noalign{\hrule height 1.5pt}

 Mercury & Venus  & Earth & Mars\\
$-0.0036\pm 0.0050$ & $0.53\pm 0.30$ & $-0.0002\pm 0.0004$ & $0.0001\pm 0.0005$\\
\hline

\noalign{\hrule height 1.5pt}
\end{tabular}

\end{table}}
Although the original purpose\footnote{The goal by Pitjeva (2005a)
was to make a test of the quality of the previously obtained
general solution in which certain values of $\beta, \gamma$ and
$J_2$, were obtained. If the construction of the ephemerides was
satisfactory, very small ``residual" effects due to such
parameters should have been found. She writes: ``At present, as a
test, we can determine [...] the corrections to the motions of the
planetary perihelia, which allows us to judge whether the values
of $\beta,\ \gamma,$ and $J_2$ used to construct the ephemerides
are valid.''. The smallness of the extra-perihelion precessions
found in her particular test-solution is interpreted by Pitjeva as
follows: ``Table 3 shows that the parameters $\beta=1,\ \gamma=1,$
and $J_2=2\times 10^{-7}$ used to construct the EPM2004
ephemerides are in excellent agreement with the observations.'' }
of the determination of such corrections was not the measurement
of the LT effect, the results of Table 3 by Pitjeva (2005a) can be
used to take first steps towards an observational corroboration of
the existence of the solar gravitomagnetic force. Indeed, the
uncertainties in the predicted values of the LT precessions
induced by the error in $L_{\odot}$ (Pijpers 1998; 2003) amount to
$1\times 10^{-5}$ \asec\ for Mercury, $7\times 10^{-7}$ \asec\ for
the Earth and $2\times 10^{-7}$ \asec\ for Mars: they are far
smaller than the errors in Table \ref{pitabl}, so that a genuine
comparison with the measured precessions make sense.

\begin{figure}
\begin{center}
\includegraphics[width=13cm,height=11cm]{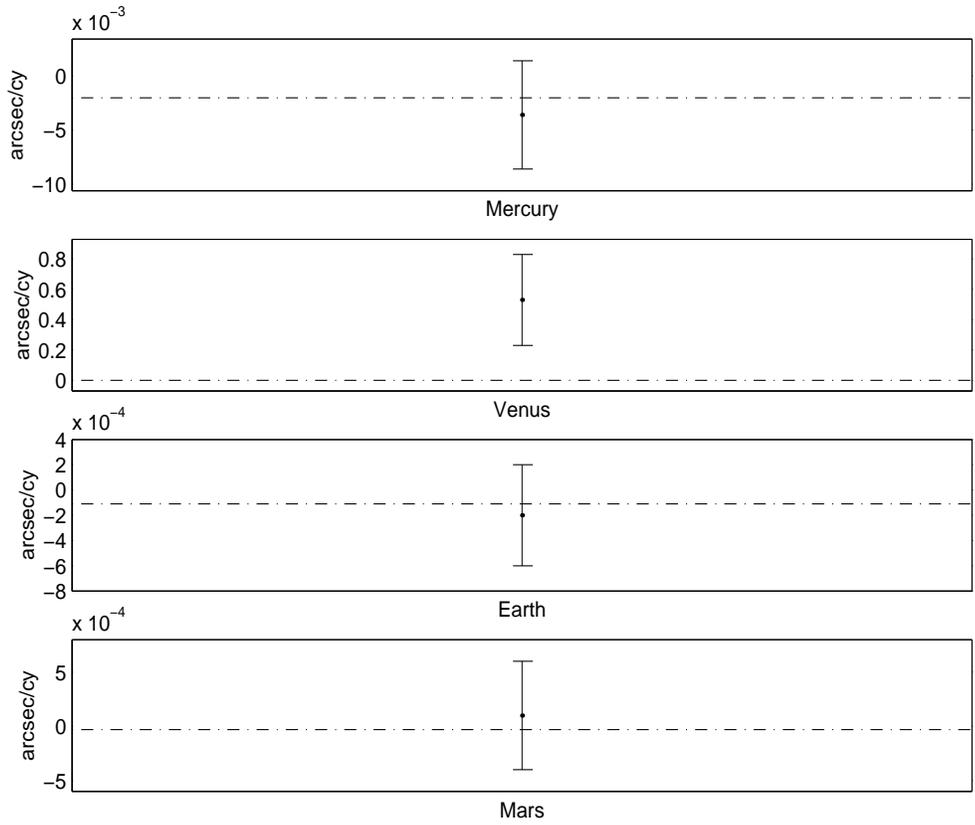}
\end{center}
\caption{\label{figura} The horizontal dash-dotted lines represent
the predicted values of the LT secular precessions of the
perihelia of Mercury, Venus, the Earth and Mars according to GTR.
The vertical solid lines represent the values  of the additional
secular precessions of Mercury, Venus, the Earth and Mars
determined by Pitjeva (2005a) along with their error bars. The
predictions of the LT effect by GTR (1 in normalized units) are
compatible with them for Mercury ($1.8\pm 2.5$), the Earth ($2\pm
4$) and Mars ($-3.3\pm 16.6$).}
\end{figure}

By comparing Table \ref{solgm} and Table \ref{pitabl} of this
paper it turns out that the
 predictions of GTR for the LT effect are compatible with the small determined
corrections to the secular motions of the planetary perihelia
for\footnote{In the case of Venus the discrepancy between the
predicted and the measured values is slightly larger than the
measurement error. For such a planet the perihelion is not a good
observable because of the small eccentricity of its orbit ($e_{\rm
Venus}=0.0066$).} Mercury ($-0.0086$\asec$ < -0.0020$\asec $<
0.0014$ \asec), the Earth ($-0.0006$\asec $< -0.0001$\asec $<
0.0002$ \asec) and Mars (-0.0004 \asec $<$ -$3\times 10^{-5}$
\asec $<$ 0.0006 \asec). In normalized units $\mu$ ($\mu_{\rm
GTR}=1$) we have $\mu_{\rm obs}^{\rm Mercury}=1.8\pm 2.5$,
$\mu_{\rm obs}^{\rm Earth}=2\pm 4$ and $\mu_{\rm obs}^{\rm
Mars}=-3.3\pm 16$. Figure \ref{figura} summarizes the obtained
results.
The discrepancies between the predicted and the determined values
are reported in Table \ref{erri} of this paper.
 They are smaller than the measurement uncertainties, so that a
 $\chi^2=\sum\left(\frac{P-D}{E}\right)^2=0.2$ can be obtained.
 It must be noted that the determined extra-precessions of Table \ref{pitabl}
 are
 also compatible with zero, but at a worse level. Indeed,
 $\chi^2=0.8$ in this case.
{\small\begin{table}\caption{ Comparison between the predicted
values (P) of the LT precessions of the perihelia of Mercury, the
Earth and Mars (Table \ref{solgm}) and the determined values (D)
of the extra-precessions of their perihelia (Table \ref{pitabl}).
Their differences are smaller than the errors (E). From them a
$\chi^2=0.2$ can be obtained. Note that the $\chi^2$ for zero is
0.8. }\label{erri}

\begin{tabular}{lll} \noalign{\hrule height 1.5pt}

Planet & P-D (\asec) & E (\asec) \\
Mercury & 0.0016 & 0.0050\\
Earth & 0.0001 & 0.0004\\
Mars & -0.0001 & 0.0005\\ \hline

\noalign{\hrule height 1.5pt}
\end{tabular}

\end{table}}

A way to improve the robustness and reliability of such a test
would be to vary the adopted values for the solar oblateness
within the currently accepted ranges and investigate the changes
in the fitted values of the extra-precessions. Moreover, it would
also be important to produce an analogous set of solutions with
$\beta,\gamma$ and $J_2^{\odot}$ fixed in which the
extra-precessions of the nodes are determined; in this way it would be possible to use only Mercury.
\subsection{Some possible systematic errors due to other competing effects}
In order to check our conclusion that the LT effect is the main
responsible for the observed secular corrections to the planetary
perihelia $\Delta\dot\varpi_{\rm obs }$ let us focus on Mercury
and on the known perturbations which could induce a secular
extra-perihelion advance due to their mismodelling.

The major sources of secular advances of the perihelia are the
Schwarzschild gravitoelectric part  of the solar gravitational field and the quadrupolar
mass moment $J_2^{\odot}$ of the Sun. Their nominal effects on the
longitudes of perihelion of the inner planets are quoted in Table
\ref{solge} and Table \ref{solj2} of this paper; the analytical
expressions are \eqia\dot\varpi_{\rm GE}&=&\rp{3nGM}{c^2
a(1-e^2)},\\
\dot\varpi_{J_{2}}&=&\rp{3}{2}\rp{n
J_2}{\left(1-e^2\right)^2}\left(\rp{R}{a}\right)^2\left(1-\rp{3}{2}\sin^2
i \right)\equiv\dot\varpi_{.2} J_2,\eqfa where $n=\sqrt{GM/a^3}$
is the Keplerian mean motion and $R$ is the mean equatorial radius
of the central body.
{\small\begin{table}\caption{ Nominal values of the secular
post-Newtonian gravitoelectric precessions of the longitudes of  perihelion
$\varpi$ of Mercury, Venus, Earth and Mars in \asec. Their
mismodelled amplitudes are fixed by the uncertainties in $\gamma$
and $\beta$ which are of the order of $0.01\%$ (Pitjeva
2005a). However, deviations from their relativistic values are expected only at a $10^{-6}-10^{-7}$ level (Damour and Nordtvedt 1993).}\label{solge}

\begin{tabular}{llll}
\noalign{\hrule height 1.5pt}

 Mercury & Venus  & Earth & Mars\\
42.9812 & 8.6247 & 3.8387 & 1.3509\\ \hline

\noalign{\hrule height 1.5pt}
\end{tabular}

\end{table}}
{\small\begin{table}\caption{ Nominal values of the classical
secular precessions of the longitudes of  perihelion $\varpi$ of
Mercury, Venus, Earth and Mars, in \asec, induced by the solar
quadrupolar mass moment $J_2^{\odot}$. The value
$J_2^{\odot}=2\times 10^{-7}$ used in (Pitjeva 2005a) has been
adopted. Their mismodelled amplitudes are fixed by the uncertainty
in $J_2^{\odot}$ which is of the order of $\sim
10\%$.}\label{solj2}

\begin{tabular}{llll}
\noalign{\hrule height 1.5pt}

 Mercury & Venus  & Earth & Mars\\
0.0254 & 0.0026 & 0.0008 & 0.0002 \\ \hline

\noalign{\hrule height 1.5pt}
\end{tabular}

\end{table}}
In view of their large size with respect to the LT effect, one
could legitimately ask if the determined extra-precessions are due
to the systematic errors in such competing secular rates. An
a-priori analytical analysis shows that it should not be the case.

\subsubsection{The impact of the solar oblateness}
In regard to $J_2^{\odot}$, which is still rather poorly known,
only values measured in such a way that no a priori `imprinting'
effects occurred should be considered for our purposes.

E.g., the most recent determinations of the solar oblateness based
on astrophysical techniques yield values close to $2.2\times
10^{-7}$ (Patern\`{o} et al. 1996; Pijpers 1998; Mecheri et al.
2004) with discrepancies between the various best estimates of the
same order of magnitude of their errors, i.e. $\sim 10^{-9}$. Let
us see if such determinations are compatible with the determined
extra-advances of perihelia. By assuming a correction of $\sim
10\%$ of the adopted reference value by Pitjeva, i,e. $\delta
J_2^{\odot}=2\times 10^{-8}$, the resulting residual precession
due to the solar oblateness would amount to $+0.0025$\asec\ for
Mercury.
It falls outside the measured range.

A way to a priori cancel out any possible impact of the
uncertainty in the solar oblateness consists in suitably combining
the perihelia advances of two planets so to de-correlate by
construction the LT effect and the precessions due to $J_2$. This
approach allows to extract the gravitomagnetic signal
independently of the solar quadrupolar mass moment. On the other
hand, it is also possible to measure a correction $\delta
J_2^{\odot}$ to it independently of the LT effect. In turn, such
value for $\delta J_2^{\odot}$ can be used to check if it can
accommodate the determined extra-precessions of Table
\ref{pitabl}. Let us write \eqi\left\{\begin{array}{lll}
\Delta\dot\varpi_{\rm obs}^{\rm Mercury}=\dot\varpi_{.2}^{\rm
Mercury}\delta J_2^{\odot}+
\dot\varpi_{\rm LT}^{\rm Mercury}\mu,\\\\
\Delta\dot\varpi_{\rm obs}^{\rm Earth}=\dot\varpi_{.2}^{\rm
Earth}\delta J_2^{\odot}+\dot\varpi_{\rm LT}^{\rm
Earth}\mu.\lb{siste}\end{array}\right.\eqf
By solving \rfr{siste} with respect to $\mu$ it is possible to
obtain \eqi\mu=\rp{\Delta\dot\varpi_{\rm obs}^{\rm
Mercury}+c_1\Delta\dot\varpi_{\rm obs}^{\rm
Earth}}{\dot\varpi_{\rm LT}^{\rm Mercury}+c_1\dot\varpi_{\rm
LT}^{\rm Earth}},\lb{sistem}\eqf with \eqi
c_1=-\rp{\dot\varpi_{.2}^{\rm Mercury}}{\dot\varpi_{.2}^{\rm
Earth}}\sim-\left(\rp{a^{\rm Earth}}{a^{\rm Mercury
}}\right)^{7/2}\left(\rp{1-e^2_{\rm Earth}}{1-e^2_{\rm Mercury
}}\right)^2=-30.1930,\lb{coff}\eqf and \eqi \dot\varpi_{\rm
LT}^{\rm Mercury}+c_1\dot\varpi_{\rm LT}^{\rm Earth}= 0.0013\
''{\rm cy }^{-1}.\eqf The combination of \rfr{sistem} is not
affected by the solar oblateness whatever its real value is:
indeed, \eqi\dot\varpi_{.2}^{\rm Mercury}+c_1\dot\varpi_{.2}^{\rm
Earth}=0,\ \forall\ J_2^{\odot}.\eqf By inserting the value of
\rfr{coff}, the figures of Table \ref{solgm} and the results of
Table \ref{pitabl} in \rfr{sistem} one obtains for such a
combination\footnote{The error has been evaluated by summing in
quadrature the errors of Table \ref{pitabl} according to
\rfr{sistem}.} $\mu_{\rm obs}=1.8\pm 10$.  As can be noted, the
best estimate for $\mu$ does not change with respect to the case
of Mercury's perihelion only, as if departures of the solar
oblateness from the adopted reference value were of little
importance. Indeed, if we solve \rfr{siste} with respect to the
correction to the Sun's quadrupolar mass moment the equation
\eqi\delta J_2^{\odot}= \rp{\Delta\dot\varpi_{\rm obs}^{\rm
Mercury}+d_1\Delta\dot\varpi_{\rm obs}^{\rm
Earth}}{\dot\varpi_{.2}^{\rm Mercury}+d_1\dot\varpi_{.2}^{\rm
Earth}},\lb{sistem2}\eqf with \eqi d_1=-\rp{\dot\varpi_{\rm LT
}^{\rm Mercury}}{\dot\varpi_{\rm LT}^{\rm
Earth}}\sim-\left(\rp{a_{\rm Earth}}{a_{\rm
Mercury}}\right)^3\left(\rp{1-e^2_{\rm Earth}}{1-e^2_{\rm Mercury
}}\right)^{3/2}=-18.3864,\eqf is obtained. \Rfr{sistem2} allows to
measure the correction to the adopted value of $J_2^{\odot}$, by
construction, independently of the LT effect in the sense that
\eqi\dot\varpi^{\rm Mercury}_{\rm LT}+d_1\dot\varpi_{\rm LT}^{\rm
Earth}=0,\ \forall\ L_{\odot}.\eqf The result is\footnote{Note
that Pitjeva (2005a) derived a negative correction $\delta
J_2^{\odot}$ of order $10^{-8}$ from the determined extra-advance
of Mercury's perihelion only, without taking into account the
biasing impact of the LT effect which amounts to $\sim 8\%$ for
Mercury, as can be inferred from Table \ref{solgm} and Table
\ref{solj2}. } \eqi\delta J_2^{\odot}=(+0.01\pm 0.47)\times
10^{-7}.\lb{correz}\eqf Such value can be considered as a
dynamical measurement of the solar oblateness independent of the
general relativistic gravitomagnetic features of motion. It
induces a ``residual'' precession of $+0.0001$ \asec\ on Mercury's
perihelion, which is smaller than its observed extra-advance and
the related error. For the Earth the ``residual'' effect of
\rfr{correz} would amount to $+4\times 10^{-6}$ \asec.

%
%

It is important to note that the combination of \rfr{sistem}
yields $\chi^2=0.007$, while, by assuming zero extra-precessions,
one has $\chi^2=0.03$: also in this case, the relativistic
prediction of LT is in better agreement with data than the
zero-effect hypothesis.
\subsubsection{The post-Newtonian gravitoelectric precessions}
Although the large nominal values of their precessions, the
post-Newtonian gravitoelectric terms do not represent a problem.
Indeed, they are fully included in the dynamical force models of
EPM2004 in terms of the PPN parameters $\beta$ and $\gamma$ which
are presently known at a $10^{-4}-10^{-5}$ level (Pitjeva 2005a;
Bertotti et al. 2003).  Moreover, theoretical deviations from the
GTR values are expected at a $10^{-6}-10^{-7}$ level (Damour and
Nordtvedt 1993).
\subsubsection{The impact of the asteroids}
As already noted, the dynamical force models adopted in EPM2004
also include the action of the major asteroids and of the ecliptic
ring which accounts for the other minor bodies. Indeed, it has
recently pointed out that their impact limits the accuracy of the
inner planets' ephemerides over time-scales of a few decades
(Standish and Fienga 2002) in view of the relatively high
uncertainty in their masses (Krasinsky et al. 2002; Pitjeva 2005b)
Recently, Fienga and Simon (2005) have shown that also Mercury's
orbit is affected to a detectable level by secular perturbations
due to the most important asteroids.

May it happen that the mismodelled part of such secular
precessions could explain the observed $\Delta\dot\varpi_{\rm
obs}^{\rm Mercury}$?

From Table 3 by Fienga and Simon (2005) the nominal amplitude of
the secular perturbations on $\varpi^{\rm Mercury}$ due to 295
major asteroids can be calculated. It turns out to be $0.0004$
\asec; even assuming a conservative $\sim 10\%$ uncertainty
(Pitjeva 2005b), it is clear that the asteroids are not the cause
of the determined extra-shift of Mercury's perihelion.
\subsubsection{The impact of non-Einsteinian effects}
In regard to other possible sources of extra-secular precessions
of the planetary perihelia outside the scheme of the
Newton-Einstein gravity, recently it has been shown by Lue and
Starkman (2003)  that the multidimensional braneworld gravity
model by Dvali, Gabadadze and Porrati (2000) predicts also a
secular perihelion shift in addition to certain cosmological
features.
By postulating that the current cosmic acceleration is entirely
caused by the late-time self-acceleration, constraints from Type
1A Supernov\ae\ data yield a value
 of $\sim 0.0005$ \asec\ for the Lue-Starkman planetary precessions. Also this effect is too
small to accommodate the determined additional perihelion advance of
Mercury.
\subsection{Analysis of other independent data}
The shift of the perihelion yields a variation of the planet's
range that can be expressed as (Nordtvedt 2000) $\Delta
r=ea\Delta\varpi$. In the case of Mercury  the centennial
variation due to the Lense-Thirring precession amounts to -115 m.

Fienga et al. (2005) used their numerical
ephemerides\footnote{They include a complete suit of improved
dynamical models, apart from just the post-Newtonian
gravitomagnetic forces whose effects are, thus, fully present in
the determined residuals.} INPOP, recently produced at the
Institute of m\'{e}canique c\'{e}leste et de calcul des
\'{e}ph\'{e}m\'{e}rides (IMCCE), to fit different kinds of
observations including also the radar-ranging data to the inner
planets. They found for the Mercury's range residual the value
$\delta r_{\rm meas}=-95.6\pm 784$ m. It is worth noting that in
obtaining these results also the impact of the asteroids on
Mercury's orbital motion was accounted for. As it can be noted,
also in this case the errors are large, but the general
relativistic prediction for the Lense-Thirring effect are in
better agreement with the  data than the hypothesis of null effect
([$({\rm P}-{\rm D})/ {\rm E}]^2=6\times 10^{-4}$ and [$({\rm
P}-{\rm D})/ {\rm E}]^2=1\times 10^{-2}$, respectively).
\section{Constraints on a Yukawa-like fifth force} The
differences between the determined extra-precessions and the
predicted LT rates of Table \ref{erri} of this paper can also be
used to strongly constrain, at planetary length-scales
$10^{10}-10^{11}$ m, departures from the inverse-square-law
phenomenologically parameterized in terms of  the magnitude
$|\alpha|$ of the strength of a Yukawa-like fifth force
(Fischbach et al. 1986; Adelberger et al. 2003). Indeed, a potential \eqi U_{\rm
Yukawa}=-\rp{GM}{r}\left[1+\alpha\exp\left(-\rp{r}{\lambda}\right)\right],\eqf
where $\lambda$ is the range of such a hypothesized force, can
produce a secular perihelion advance over scales $\lambda$
comparable to $a$ (Lucchesi 2003) \eqi \dot\varpi_{\rm
Yukawa}\propto \rp{\alpha n}{e}.\eqf By using the figures of Table \ref{erri} it is possible to constrain
$\alpha$ to $\approx 10^{-12}-10^{-13}$ level at $r=\lambda\approx 1$ A.U.
The most recently published constraints in the planetary range are
at $10^{-9}- 10^{-10}$ level (Bertolami and Paramos 2005; Reynaud
and Jaekel 2005).
\section{Discussion and
conclusions} In this paper we discussed the possibility of
performing new tests of post-Newtonian gravity in the Solar
System. To this aim, we analyzed the estimated corrections to the secular
rates of the perihelia of the inner planets of the Solar System
recently determined by E.V. Pitjeva (Institute of Applied Astronomy, Russian Academy of Sciences).
She used the EPM2004
ephemerides with a wide range of observational data spanning
almost one century; in a particular solution, she solved also for
the secular motions of the perihelia by keeping fixed the PPN
parameters $\beta$ and $\gamma$ and the solar quadrupole mass
moment $J_2^{\odot}$, and neglecting the gravitomagnetic force in
the dynamical force models.

It turns out that the post-Newtonian LT secular precessions
predicted by GTR are compatible with the determined
extra-precessions for Mercury, the Earth and, to a lesser extent,
Mars: in normalized units ($\mu=1$ in GTR) we have $\mu_{\rm
obs}=1.8\pm 2.5$ for Mercury, $\mu_{\rm obs}=2\pm 4$ for the Earth
and $\mu_{\rm obs}=-3.3\pm 16.6$ for Mars. A suitable combination
of the perihelia of Mercury and the Earth, which cancels out any
possible bias by $J_2^{\odot}$, yields $\mu_{\rm obs}=1.8\pm 10$.
It must be noted that the errors are still large and the data are
compatible also with the hypothesis of zero extra-precessions, but
at a worse level with respect to the relativistic LT prediction.
If confirmed by further, more extensive and robust data analysis
by determining, e.g., the extra-precessions of the nodes as well,
it would be the first observational evidence of the solar
gravitomagnetic field. The processing of further amounts of data,
along with  those expected in future from the forthcoming planetary
mission BepiColombo and, perhaps,  Messenger and Venus Express as well, although to a lesser extent,
will further
improve the accuracy in determining the orbital motion of these
planets and, consequently, the precision of the LT tests.

A by-product of the present analysis is represented by new, strong
constraints ($10^{-12}-10^{-13}$) on the strength of a Yukawa-like
fifth force at scales of about one Astronomical Unit.


\section*{Acknowledgements}
I thank E.V. Pitjeva for helpful clarifications about
her measured extra-precessions, J.-F. Pascual-S$\acute{\rm
a}$nchez, O. Bertolami and G. Melki for useful comments and
references. I am grateful to the anonymous referees whose comments greatly improved
the paper.



\begin{thebibliography}{xxxxx}

\item []
Adelberger, E.G., Heckel, B.R., Nelson, A.E., 2003. Tests of the
gravitational inverse-square law. Annual Review of Nuclear and
Particle Science 53(December 2003), 77-121.

\item []
Ashby, N., Allison, T., 1993. Canonical planetary equations for
velocity-dependent forces, and the Lense-Thirring precession.
Celestial Mechanics and Dynamcal Astronomy  57(4), 537-585.



\item []
Barker, B.M., O'Connell, R.F., 1974. Effect of the rotation of the
central body on the orbit of a satellite. Physical Review D 10(4),
1340-1342.



\item []
Bertolami, O., Paramos, J., 2005. Astrophysical Constraints on
Scalar Fields Models. Physical Review D 71(2), 023521.

\item []
Bertotti, B., Iess, L., Tortora, L., 2003. A test of general
relativity using radio links with the Cassini spacecraft. Nature
425(25 September 2003), 374-376.

\item [] Ciufolini, I.,  Wheeler, J.A., 1995. Gravitation and
Inertia. Princeton University Press, Princeton.

\item []
Ciufolini, I., Lucchesi, D.M., Vespe, F., and Mandiello, A., 1996. Measurement of Dragging of Inertial Frames and Gravitomagnetic Field Using Laser-Ranged Satellites. Il Nuovo Cimento A 109(5) 575-590.


\item []
Ciufolini, I., Pavlis, E.C., 2004. A confirmation of the general
relativistic prediction of the Lense–-Thirring effect. Nature
431(21 October 2004), 958-960.



\item []
Ciufolini, I., Pavlis, E.C., 2005. On the measurement of the
Lense-Thirring effect using the nodes of the LAGEOS satellites, in
reply to ``On the reliability of the so far performed tests for
measuring the Lense-Thirring effect with the LAGEOS satellites''
by L. Iorio. New Astronomy 10(8), 636-651.






\item []
Cugusi, L., Proverbio, E., 1978. Relativistic Effects on the
Motion of Earth's Artificial Satellites. Astronomy and
Astrophysics 69, 321-325.


\item []
Damour, T., Nordtvedt, K., 1993. General relativity as a
cosmological attractor of tensor-scalar theories. Physical Review
Letters 70(15), 2217-2219.

\item []
de Sitter, W., 1916. On Einstein's Theory of Gravitation and its
Astronomical Consequences. Monthly Notices of the Royal
Astronomical Society 76(9), 699-728.


\item [] Dvali, G., Gabadadze, G.,  Porrati, M., 2000.
4D Gravity on a Brane in 5D Minkowski Space. Physics Letters B
485(1-3), 208-214.


\item []
Everitt, C.W.F, 1974. The Gyroscope Experiment I. General
Description and Analysis of Gyroscope Performance. In: Bertotti,
B. (Ed.), Proc. Int. School Phys. "Enrico Fermi" Course LVI. New
Academic Press, New York,  pp. 331-360. Reprinted in: Ruffini,
R.J., Sigismondi, C. (Eds.), 2003. Nonlinear Gravitodynamics.
World Scientific, Singapore, pp. 439-468.


\item []
Everitt, C.W.F.,  et al., 2001. Gravity Probe B: Countdown to
Launch. In: L\"{a}mmerzahl, C., Everitt, C.W.F., Hehl, F.W.
(Eds.), Gyros, Clocks, Interferometers...: Testing Relativistic
Gravity in Space. Springer, Berlin,  pp. 52-82.



\item []
Fienga, A.,  Simon, J.-L., 2005. Analytical and numerical studies
of asteroid perturbations on solar system planet dynamics.
Astronomy and Astrophysics 429(2), 361-367.

\item []
Fischbach, E., Sudarsky, D., Szafer, A., Talmadge, C., and Aronson, S. H., 1986.
Reanalysis of the E\"{o}tv\"{o}s experiment, Physical Review Letters 56(1), 3-6.

\item []
Fitch, V.L., et al., 1995. Review of Gravity Probe B. National Academic Press, Washington DC.


\item [] Fienga, A., Laskar, J., Simon, J.L., Manche, H.,
Gastineau, M., 2005. IMCCE Planetary Ephemerides: Present and
Future, In: Turon, C.,  O'Flaherty, K.S.,  Perryman, M.A.C. (Eds.)
{Proceedings of the Gaia Symposium: The Three-Dimensional Universe
with Gaia, held at Observatoire de Paris-Meudon, 4-7 October
2004}, (ESA SP-576). From
http://www.rssd.esa.int/index.php?project=Gaia$\&$page=Gaia$\_$2004$\_$Proceedings,
supercedes printed version, pp. 293-296.


\item []
Iorio, L., 2001. An alternative derivation of the Lense-Thirring
drag on the orbit of a test body. Il Nuovo Cimento B 116(7),
777-789.




\item []
Iorio, L., 2005a. On the reliability of the so far performed tests
for measuring the Lense-Thirring effect with the LAGEOS
satellites.  New Astronomy 10(8), 603-615.


\item []
Iorio, L., 2005b. The impact of the new Earth gravity models on
the measurement of the Lense-Thirring effect with a new satellite.
New Astronomy 10(8), 616-635.


\item []
Iorio, L., 2005c. Is it possible to measure the Lense-Thirring
effect on the orbits of the planets in the gravitational field of
the Sun?. Astronomy and Astrophysics 431(1), 385-389.



\item [] Iorio, L., 2007. An assessment of the measurement of
the Lense–Thirring effect in the Earth gravity field, in reply to:
"On the measurement of the Lense–Thirring effect using the nodes
of the LAGEOS satellites, in reply to "On the reliability of the
so far performed tests for measuring the Lense–Thirring effect
with the LAGEOS satellites" by L. Iorio," by I. Ciufolini and E.
Pavlis, Planetary Space Science 55(4), 503-511.

\item []
Iorio, L., 2006a. A critical analysis of a recent test of the
Lense-Thirring effect with the LAGEOS satellites. Journal of
Geodesy 80(3), 128-136. (Preprint
http://www.arxiv.org/abs/gr-qc/0412057v8).


\item []
Iorio, L., 2006b. Evidence of the gravitomagnetic field of Mars.
Classical and Quantum Gravity 23(17), 5451-5454.



\item []
Iorio, L.,  Morea, A., 2004. The impact of the new Earth gravity
models on the measurement of the Lense-Thirring effect. General
Relativity and  Gravitation 36(6), 1321-1333. (Preprint
http://www.arxiv.org/abs/gr-qc/0304011).




\item []
Kopeikin, S.M., 2007. Comment on ``The gravitomagnetic influence on gyroscopes and on the lunar orbit", gr-qc/0702120.

\item []
Krasinsky, G.A., Pitjeva, E.V., Vasiljev, M.V., Yagudina, E.I.,
2002. Hidden  Mass in the Asteroid Belt. Icarus 158(1), 98-105.



\item []
Lense, J.,  Thirring, H., 1918. \"{U}ber den Einfluss der
Eigenrotation der Zentralk{\"{o}}rper auf die Bewegung der
Planeten und Monde nach der Einsteinschen Gravitationstheorie.
{Phys. Z.} {19}, 156-163. Translated and discussed by Mashhoon,
B., Hehl F.W., Theiss, D.S., 1984. On the Gravitational Effects of
Rotating Masses: The Thirring-Lense Papers. General Relativity and
Gravitation 16(8), 711-750. Reprinted in: Ruffini, R.J.,
Sigismondi, C. (Eds.), 2003. Nonlinear Gravitodynamics. World
Scientific, Singapore, pp. 349-388.


\item [] Lucchesi, D.M., 2003. LAGEOS II perigee shift and
Schwarzschild gravitoelectric field. Physics Letters A 318(3),
234-240.

\item []
Lucchesi, D.M., 2005. The impact of the even zonal harmonics
secular variations on the Lense-Thirring effect measurement with
the two Lageos satellites. International Journal of Modern Physics
D 14(12), 1989-2023.


\item []
Lue, A.,  Starkman, G., 2003. Gravitational Leakage into Extra
Dimensions Probing Dark Energy Using Local Gravity. Physical
Review D 67(6), 064002.



\item []
Mashhoon, B., Gronwald, F., Lichtenegger, H., 2001.
Gravitomagnetism and the Clock Effect. In: L\"{a}mmerzahl C,
Everitt, C.W.F., and Hehl, F.W. (Eds.), Gyros, Clocks,
Interferometers...: Testing Relativistic Gravity in Space.
Springer, Berlin, pp. 83-108.




\item []
Mecheri, R., Abdelatif, T., Irbah, A., Provost, J.,  Berthomieu,
G., 2004.  New values of gravitational moments J2 and J4 deduced
from helioseismology. Solar Physics 222(2), 191-197.

\item []
Muhlfelder, B., Mac Keiser, G., Turneaure, J., 2007. Gravity Probe B Experiment Error, poster L1.00027 presented at the American Physical Society (APS) meeting in Jacksonville, Florida, on 14-17 April 2007.


\item []
Murphy, T.W., Nordtvedt K., and Turyshev, S.G., 2007. Gravitomagnetic Influence on Gyroscopes and on the Lunar Orbit.
Physical Review Letters 98(7), 071102.



\item [] Newhall, X.X., Standish, E.M.,  Williams, J.G., 1983.
DE 102-A numerically integrated ephemeris of the moon and planets
spanning forty-four centuries. Astronomy and Astrophysics 125(1),
150-167.


\item []
Nordtevdt, K., 1988. Gravitomagnetic interaction and laser ranging
to Earth satellites. Physical Review Letters 61(23), 2647-2649.


\item [] Nordtvedt, K., 2000. Improving gravity theory tests with
solar system ``grand fits''. Physical Review D 61(12), 122001.

\item []
Nordtvedt, K., 2003. Some considerations on the varieties of frame
dragging. In: Ruffini, R.J., Sigismondi, C. (Eds.),  Nonlinear
Gravitodynamics. The Lense--Thirring Effect.   World Scientific,
Singapore,  pp. 35-45.






\item [] Patern\`{o}, L., Sofia, S.,  Di Mauro, M.P., 1996. The
rotation of the Sun's core. Astronomy and Astrophysics 314,
940-946.


\item []
Pavlis, E.C., 2002. Geodetic contributions to gravitational experiments in space. In:
Cianci, R., Collina, R., Francaviglia, M., Fr\'{e}, P. (Eds.),
Recent Developments in General Relativity. 14th SIGRAV Conference on General Relativity and Gravitational Physics, Genova, Italy, September 18-22, 2000. Springer, Milano, pp. 217-233.


\item []
Pijpers, F.P., 1998. Helioseismic determination of the solar
gravitational quadruople moment. Monthly Notices of the Royal
Astronomical Society 297(3) L76-L80.

\item [] Pijpers, F.P., 2003. Asteroseismic determination of stellar
angular momentum. Astronomy and Astrophysics 402(2), 683-692.

\item []
Pitjeva, E.V., 2005a. Relativistic Effects and Solar Oblateness
from Radar Observations of Planets and Spacecraft. Astronomy
Letters 31(5), 340-349.

\item []
Pitjeva, E.V., 2005b. High-Precision Ephemrides of Planets-EPM and
Determinations of Some Astronomical Constants. Solar System
Research 39(3), 176-186.


\item []
Pugh, G.E., 1959. Proposal for a Satellite Test of the Coriolis
Prediction of General Relativity, WSEG, Research Memorandum No.
11, reprinted in:   Ruffini, R.J., Sigismondi, C. (Eds.), {
Nonlinear Gravitodynamics. The Lense--Thirring Effect}.   World
Scientific, Singapore, pp. 414-426.


\item [] Reynaud, S., Jaekel, M.-T., 2005. Testing the Newton Law
at Long Distances. Internationa Journal of Modern Physics A
20(11), 2294-2303.


\item []
Ries, J.C., Eanes, R.J., Watkins, M.M., Tapley, B., 1989. Joint
NASA/ASI Study on Measuring the Lense-Thirring Precession Using a
Second LAGEOS Satellite. CSR-89-3, Center for Space Research, The
University of Texas at Austin.


\item []
Ries, J.C., Eanes, R.J., Tapley, B.D., 2003a. Lense-Thirring
Precession Determination from Laser Ranging to Artificial
Satellites. In: Ruffini, R.J., Sigismondi, C. (Eds.), {Nonlinear
Gravitodynamics. The Lense--Thirring Effect}, World Scientific,
Singapore, pp. 201-211.

\item []
Ries, J.C., Eanes, R.J., Tapley, B.D., Peterson, G.E., 2003b.
Prospects for an Improved Lense-Thirring Test with SLR and the
GRACE Gravity Mission. In: Noomen, R., Klosko, S., Noll, C.,
Pearlman, M. (Eds.), {Proceedings of the 13th International Laser
Ranging Workshop, NASA CP 2003-212248}.  NASA Goddard, Greenbelt.
(Preprint
http://cddisa.gsfc.nasa.gov/lw13/lw$\_${proceedings}.html$\#$science).




\item []
Ruggiero, M.L., Tartaglia, A., 2002. Gravitomagnetic Effects. Il
Nuovo Cimento B 117(7), 743-767.

\item []
Schiff, L., 1960. On Experimental Tests of the General Theory of
Relativity. American Journal of Physics 28(4), 340-343.

\item []
Sereno, M., 2003. Gravitational lensing by stars with angular
momentum. Monthly Notices of the  Royal Astronomical Society
344(3), 942-950.

\item []
Sereno, M., 2005a. Detecting gravitomagnetism with rotation of
polarization by a gravitational lens.  Monthly Notices of the
Royal Astronomical Society 356(1), 381-385.

\item []
Sereno, M., 2005b. On gravitomagnetic time-delay by extended
lenses. Monthly Notices of the  Royal Astronomical Society
357(4), 1205-1210.




\item []
Soffel, M.H., 1989. Relativity in Astrometry, Celestial Mechanics
and Geodesy. Springer, Berlin.



\item [] Standish, E.M.,  Fienga, A., 2002. Accuracy limit of modern
ephemerides imposed by the uncertainties in asteroid masses.
Astronomy and Astrophysics  384(1), 322-328.

\item []
Stella, L. Cui, W., Chen, W., Zhang, S.N., Van Der Klis, M.,
Karas, V., Semer\'{a}k, O., De Felice, F., Dov\v{c}iak, M.,
Casini, H., Montemayor, R., Morsink, S.M., Silbergleit, A.S.,
Wagoner, R.V., Khanna, R., Markovic, D., Lamb, F.K., 2003. Section
E: Probing the Gravitomagnetic Lense-Thirring effect with Neutron
Stars and Black Holes. In: Ruffini, R.J., Sigismondi, C. (Eds.),
{Nonlinear Gravitodynamics. The Lense--Thirring Effect}. World
Scientific, Singapore, pp. 235-345.

\item []
Thorne, K., Price, R.H., and MacDonald, D.A., 1986.
Black Holes, the Membrane Paradigm. Yale University Press, New Haven and London.

\item [] Will, C M. 1993, Theory and Experiment in Gravitational
Physics, 2nd edition. Cambridge University Press, Cambridge.


\end{thebibliography}
\end{document}